\documentclass[12pt]{article}

\usepackage{lmodern} 
\usepackage{amsmath,amsfonts,amssymb}
\usepackage{amsthm}   
\usepackage{latexsym}
\usepackage{natbib} 
\usepackage[nottoc]{tocbibind} 
\usepackage{algorithm2e}
\usepackage{microtype}
\usepackage{subcaption} 
\usepackage{multirow} 
\usepackage{booktabs} 
\usepackage{url}
\usepackage{afterpage,float,caption} 
\usepackage{rotating}   
\usepackage{graphicx}  
\usepackage{float}   
\usepackage{subcaption}
\usepackage{tablefootnote}
\usepackage{booktabs, caption, makecell} 
\usepackage{threeparttable} 

\makeatletter
\newcommand{\distas}[1]{\mathbin{\overset{#1}{\kern\z@\sim}}}%
\newsavebox{\mybox}\newsavebox{\mysim}
\newcommand{\distras}[1]{%
	\savebox{\mybox}{\hbox{\kern3pt$\scriptstyle#1$\kern3pt}}%
	\savebox{\mysim}{\hbox{$\sim$}}%
	\mathbin{\overset{#1}{\kern\z@\resizebox{\wd\mybox}{\ht\mysim}{$\sim$}}}%
}
\makeatother

\newtheorem{remark}{Remark}

\usepackage[T1]{fontenc}
\usepackage[utf8]{inputenc}
\usepackage[affil-it]{authblk}

\begin{document}

\title{\textbf{Conjugate generalized linear mixed models for clustered data}}
\author[1 2]{Jarod Y.L. Lee}
\author[1 3]{Peter J. Green}
\author[1 2 4]{Louise M. Ryan}
\affil[1]{School of Mathematical and Physical Sciences, University of Technology Sydney, Australia.}
\affil[2]{Australian Research Council Centre of Excellence for Mathematical \& Statistical Frontiers, The University of Melbourne, Australia.}
\affil[3]{School of Mathematics, University of Bristol, U.K.}
\affil[4]{Department of Biostatistics, Harvard T.H. Chan School of Public Health, U.S.}
\date{\today}
\maketitle


\begin{abstract}
	This article concerns a class of generalized linear mixed models for clustered data, where the random effects are mapped uniquely onto the grouping structure and are independent between groups. We derive necessary and sufficient conditions that enable the marginal likelihood of such class of models to be expressed in closed-form. Illustrations are provided using the Gaussian, Poisson, binomial and gamma distributions. These models are unified under a single umbrella of conjugate generalized linear mixed models, where ``conjugate'' refers to the fact that the marginal likelihood can be expressed in closed-form, rather than implying inference via the Bayesian paradigm. Having an explicit marginal likelihood means that these models are more computationally convenient, which can be important in big data contexts. Except for the binomial distribution, these models are able to achieve simultaneous conjugacy, and thus able to accommodate both unit and group level covariates.
	

\end{abstract} 

\section{Introduction}
Generalized linear mixed models \cite[]{Jiang2007,Stroup2013,Wu2010} are a broad class of models that can account for the dependency structure inherent within multilevel and longitudinal data, where the responses of units within a group are correlated. The grouping structure can be hospital, postal area, school, individual etc., and the goal is to model the response as a function of unit and group level covariates while accounting for group to group variability. For example, outcomes of patients within the same hospital are likely to be dependent due to similar risk profiles and a common clinical management practice. Generalized linear mixed models provide a natural framework for modelling dependencies by allowing for random group specific effects. \\

Despite being popular for application in areas such as marketing, biological and social sciences, generalized linear mixed models are computationally intensive to fit, especially for large scale applications such as recommender systems \cite[]{Perry2017} and discrete choice modelling \cite[]{Hensher2015,Train2009}. Inference for generalized linear mixed models is typically likelihood based, involving a multidimensional integral which usually does not have an analytic expression. Common estimation procedures include ``exact'' methods such as numerical quadrature \cite[]{Rabe-Hesketh2002} and Monte Carlo methods; approximate methods such as Laplace approximation \cite[]{Tierney1986} and penalized quasi-likelihood \cite[]{Breslow1993}; hierarchical likelihood \cite[]{Lee1996}; simulated maximum likelihood \cite[p.238--239]{Train2009}. Some of these approaches apply an expectation-maximization algorithm that treats the random effects as missing data \cite[]{McCulloch1997}. ``Exact'' methods can approximate the likelihood with arbitrary accuracy but are computational expensive. Approximate methods avoid the intractable integrals but may result in non-negligible bias \cite[]{Lin1996bias}. \\

For large scale applications, it is important that models can be fit in a reasonable time. Several methods have been proposed in various settings. \cite{Zhang2007} exploit the sparsity of predictors to achieve speedup for Bayesian hierarchical models. \cite{Luts2014} use variational approximations to fit real-time Bayesian hierarchical models to streaming data. \cite{Perry2017} proposes a moment-based procedure that is non-iterative. \cite{Scott2016} propose a fitting strategy based on the divide and recombine principle, where data are partitioned into manageable subsets and the intended statistical analysis are performed independently on each subsets before combining the results. The methods proposed by \cite{Perry2017} and \cite{Scott2016} are well suited for implementation in the context of distributed computing.  \\

In this article, we are concerned with a class of generalized linear mixed models for two-level data, where the random effects are mapped uniquely onto the grouping structure and are independent between groups. We derive necessary and sufficient conditions that enable the marginal likelihood of such class of models to be expressed in closed-form. Having an explicit marginal likelihood means that one can proceed directly to maximization without having to resort to approximate inference. We consider the most common distribution families, that is, Gaussian, Poisson, binomial and gamma. These models are unified under a single umbrella of \textit{conjugate generalized linear mixed models}, where ``conjugate'' in this context refers to the tractable form of the marginal likelihood, rather than implying inference via the Bayesian paradigm. 

\section{Exponential Family and Conjugate Prior}

The likelihood of a one parameter exponential family with dispersion can be written in the general form:
\begin{equation} 
\label{eqn:expofam}
f_{Y|\theta}(y|\theta,\phi) = \exp \left\lbrace (y\theta-b(\theta))/ \phi +c(y,\phi) \right\rbrace,
\end{equation}
\noindent for some specified functions $b(\theta)$ and $c(y,\phi)$, where $\theta$ is the \textit{canonical parameter} and can be expressed as a function of the mean $\theta(\mu)$, and $\phi$ is the \textit{dispersion parameter}, assumed known. \\

For such an exponential family, there exists a family of prior distributions on $\theta$ such that the posterior is in the same family as the prior. Such a conjugate prior for $\theta$ is defined as:
\begin{equation}
\label{eqn:conjprior}
f_{\Theta}(\theta|\chi,\nu) = g(\chi,\nu) \exp \lbrace \chi\theta  - \nu b(\theta) \rbrace,
\end{equation}
\noindent where $\chi$ and $\nu$ are parameters and $g(\chi,\nu)$ denotes the normalizing factor. \\

The posterior can be obtained by multiplying the likelihood and the prior (up to a constant of proportionality):
\begin{equation} 
\label{eqn:posterior}
f_{\Theta|y}(\theta|y,\chi,\nu,\phi) \propto \exp\lbrace c(y,\phi)\rbrace g(\chi,\nu) \exp \lbrace \theta (\chi + y/\phi) - b(\theta)(\nu+1/\phi) \rbrace,
\end{equation}
\noindent which has the same kernel as the prior, but with different parameters. The updated parameters, based on a single observation $y$, are
\begin{equation*}
\tilde{\chi} = \chi +  y/\phi ,\quad \tilde{\nu} = \nu + 1/\phi.
\end{equation*}\

For $n$ independent and identically distributed observations $y_{j}$, $j = 1,\dots,n$, it is straightforward to show that conjugacy still holds and the updated parameters are
\begin{equation*}
\tilde{\chi} = \chi +  \sum_{j} y_{j}/\phi ,\quad \tilde{\nu} = \nu + n/\phi.
\end{equation*}\

These are the standard results for independent and identically distributed data in the Bayesian context. In this article, we aim to achieve explicit marginal likelihood for generalized linear mixed models in the frequentist setting. This is attained by establishing a connection between the posterior in the Bayesian paradigm and the marginal likelihood in the frequentist paradigm, and relaxing the assumption of identical distribution. The result is a class of models where unit level covariates can be conveniently incorporated while maintaining a closed-form representation of the marginal likelihood, which we refer to  as \textit{conjugate generalized linear mixed models}.

\section{Conjugate Generalized Linear Mixed Models}

\subsection{From Bayesian formalism to frequentist inference - group level models}
We now make a transition from the Bayesian paradigm, where $\theta$ is a parameter and its distribution is the prior, to the frequentist paradigm, where $\theta$ is a group specific random effect and its distribution describes the variation between groups. \\

Specifically, consider the two-level setting where the responses $y_{ij}, j=1,\dots,n_i$ are grouped within a higher level structure indexed by $i=1,\dots,I$, with $n = n_1+\dots+n_I$ being the total number of observations across all groups. The responses are assumed to come from the same exponential family. \textit{Random effects} with a specified distribution are introduced at the group level to account for the correlation between units in a given group. Within each group, the responses are conditionally independent given the group specific random effects. Such data structure is common in many scenarios, for instance students within schools, patients within hospitals, residents within postal areas and repeated measurements from individuals. \\

For this model setup, the distribution from which the responses are drawn is governed by a group specific parameter $\theta_i$ which itself is drawn from a distribution chosen so that the resulting marginal likelihood is explicit. The marginal likelihood, obtained by integrating out the random effects, is given by
\begin{equation*}
L =  \prod_i \int \prod_j f_{Y|\theta_i}(y_{ij}|\theta_{i},\phi) f_{\Theta_i}(\theta_{i}|\chi,\nu) d\theta_{i},
\end{equation*}
where the integrand is proportional to the posterior in (\ref{eqn:posterior}). \\

Imposing a conjugate prior distribution on the random effects would ensure that the integrand comes from a recognizable density function, which would enable the marginal likelihood to be expressed in closed-form. Solving for the integral, the likelihood contribution for the entire data is
\begin{equation}
\label{eqn:marginalLik_multipleunits}
L =   \prod_i \left\lbrace \frac{\exp\left(\sum_j c(y_{ij},\phi)\right) g(\chi,\nu) }{g\left(\chi + \sum_j y_{ij} /\phi, \nu +  n /\phi \right)} \right\rbrace.
\end{equation} \

This is the formulation for group level models in the absence of of unit level covariates.  Although the random effects $\theta_{i} = \theta(\mu_{i})$ are typically expressed in terms of a monotonic transformation of $\mu_i$, interest usually lies in the distribution of $\mu_{i}$. \cite{Consonni1992conjugate} and \cite{Gutierrez1995conjugate} showed that the conjugate distribution on $\mu_i$ coincides with the prior on $\mu_i$ induced by the conjugate distribution on $\theta_i$ if and only if the exponential family has a quadratic variance function. This holds for some of the most widely used distribution, including the Gaussian, Poisson, binomial and gamma \cite[]{Morris1983natural}, providing a convenient way to incorporate group level variables, for example, via the mean $\mu_i$ using a monotonic link function.

\subsection{Relaxing the assumption of identical distribution - unit level models}
Relaxing the assumption of identical distribution, we consider the regression setting where each observation $y_{ij}$ is allowed to have a separate parameter $\theta_{ij} = \theta(x_{ij})$ that is a function of the covariates, while $\phi$, if present, is constant across all observations. We want to explore the most generic formulation that leads to marginal likelihood simplification using the idea of Bayesian conjugacy, and thus we leave open the functional dependence of $\theta_{ij}$ on $x_{ij}$ at this stage.  \\

Denote $\theta_0 = \theta(x_0)$ as the baseline parameter where $x_0$ is 
an arbitrary baseline covariate value. In this article, we assume $x_0 = 0$, but user can take any baseline appropriate for the problem at hand. Technically, $\theta_0$ is also indexed by $i$ to reflect the group correlated data structure, but this can be suppressed without ambiguity. Likewise, for ease of notation, the $i$ and $j$ indexing are suppressed for most of the remaining article. \\

\begin{remark}
	With this formulation, within a group, we can think of units with covariate configuration that deviate from the baseline characteristics as modifying $\theta_0$. This is as opposed to the standard formulation of  generalized linear mixed models, where for a given unit with a particular covariate configuration, it is the group membership that modifies the linear predictor. 
\end{remark} \

Imposing a conjugate prior distribution on $\theta_0$, the integrand of the marginal likelihood for a single observation has the form 	
\begin{equation}
\label{eqn:posterior_indiv}
\exp \left[ \lbrace y_{} \theta(x_{})  - b(\theta(x_{}))\rbrace / \phi + \chi \theta_0 - \nu b(\theta_0) \right].
\end{equation} \

\begin{remark}
	The conjugate prior distribution is placed on $\theta_0 = \theta(x_0)$, rather than explicitly on each $\theta_{ij} = \theta(x_{ij})$.
\end{remark}\

Equation (\ref{eqn:posterior_indiv}) lies in the same family as (\ref{eqn:conjprior}) in its dependence on $\theta_0$ if and only if both $\theta(x_{})$ and $b(\theta(x_{}))$ are affine functions  of $\theta_0$ and $b(\theta_0)$, i.e. if there exist functions $p$, $q$, $r$, $s$, $t$ and $u$ of $x$ such that 
\begin{gather}
\theta(x_{}) = p(x_{}) \theta_0 + q(x_{})b(\theta_0) + r(x_{}) \label{eqn:condition1}\\ 
b(\theta(x_{})) = s(x_{}) \theta_0 + t(x_{})b(\theta_0) + u(x_{}). \label{eqn:condition2}
\end{gather}\

We are interested in families where $\theta(x)$ has non-trivial dependence on $x$, that is, at least one of $p$, $q$ or $r$ must depend on $x$. When this occurs, the induced prior for $\theta(x)$ exhibits simultaneous conjugacy across all values of $x$, and the resulting model is capable of incorporating unit level covariates while maintaining a closed-form likelihood. Otherwise, this formulation reduces back to a group level model. \\

\begin{remark}
	Since $\theta_0 = \theta(x_0)$ and $b(\theta_0) = b(\theta(x_0))$, it is clear that $p(x_0) = 1$, $q(x_0) = 0$, $r(x_0) = 0$, $s(x_0) = 0$, $t(x_0) = 1$ and $u(x_0) = 0$. These constraints need to be satisfied when choosing the functional solutions for $p$, $q$, $r$, $s$, $t$ and $u$.
\end{remark}\

Conditions (\ref{eqn:condition1}) and (\ref{eqn:condition2}) can be combined to obtain
\begin{equation}
\label{eqn:condition}
b\left\lbrace p(x_{}) \theta_0 + q(x_{}) b(\theta_0) + r(x_{}) \right\rbrace = s(x_{}) \theta_0 + t(x_{})b(\theta_0) + u(x_{}).
\end{equation}\

This is the key equation in deriving the functional solutions for $p$, $q$, $r$, $s$, $t$ and $u$. Under Condition (\ref{eqn:condition}), the integrand of the marginal likelihood for a single observation is 
\begin{equation*} 
\label{eqn:posterior_CGLMM}
\exp \left[ \theta_0 \lbrace \chi +  (yp(x) - s(x)) /\phi\rbrace - b(\theta_0)\lbrace\nu + (t(x)-yq(x))/\phi \rbrace \right].
\end{equation*}\

Solving for this integral, the likelihood contribution for the observations within a single group is
\begin{equation}
\label{eqn:marginalLik_indivCov_multipleunits_singlegroup}
L =  \frac{\exp\left(\sum_j c(y_{ij},\phi)\right) g(\chi,\nu) \exp\left(  \sum_j(r(x_{ij}) y_{ij} - u(x_{ij}))/\phi \right)}{g\left(\chi + \sum_j\left(y_{ij} p(x_{ij})  - s(x_{ij})\right)/\phi, \nu +  \sum_j\left(t(x_{ij}) - y_{ij} q(x_{ij}) \right) /\phi \right)}.
\end{equation} \

For multiple groups, the likelihood contribution can be obtained by multiplying (\ref{eqn:marginalLik_indivCov_multipleunits_singlegroup}) across the group index $i$.

\section{Examples} \label{sec:examples}
\subsection{Gaussian (with known variance)}
The Gaussian density function (with known variance $\sigma^2 \geq 0$) can be written in the form
\begin{equation*}
\exp\left\lbrace \frac{y\mu_0-\mu_0^2/2}{\sigma^2} - \log\left(\sigma\sqrt{2\pi}\right) - \frac{y^2}{2\sigma^2}  \right\rbrace,
\end{equation*}
\noindent where $\mu_0 \in \mathbb{R}$ is the mean of $y$. This can be written in the form of (\ref{eqn:expofam}) if we write $\theta_0 = \mu_0$, $b(\theta_0) = \theta_0^2/2$, $\phi = \sigma^2$ and $c(y,\phi) = -\lbrace \log(2\pi\phi) + y^2/\phi \rbrace / 2$. \\

To determine the conjugate distribution for $\theta_0$, we compute the normalization factor
\begin{equation*}
g(\chi,\nu) = \left\lbrace \int \exp\left(\chi\theta_0 - \nu \frac{1}{2} \theta_0^2\right) d\theta_0  \right\rbrace ^{-1} = \sqrt{\frac{\nu}{2\pi}} \exp\left\lbrace - \left( \frac{\chi^2}{2\nu} \right) \right\rbrace,
\end{equation*}
\noindent where the integrand is the kernel of a Gaussian density function with mean E$(\theta_0) = \lambda = \chi/\nu$ and variance Var$(\theta_0) = \kappa^2 = 1/\nu$. This implies  
\begin{equation*}
\mu_0 = \theta_0 \sim \text{Gaussian} \left( \lambda , \kappa^2 \right).
\end{equation*}\

Group level covariates can be incorporated via the mean of $\mu_0$, by replacing $\lambda$ with $\lambda_i = x_i^T \beta$ for example. To incorporate unit level covariates, (\ref{eqn:condition}) requires
\begin{equation*}
b(\theta(x)) = \frac{1}{2}\left\lbrace  p(x)\theta_0 + q(x) \frac{1}{2} \theta_0^2 + r(x) \right\rbrace ^2 \equiv s(x) \theta_0 + t(x) \frac{1}{2} \theta_0^2 + u(x),
\end{equation*}
which gives the following solution set:
\begin{gather*}
p(x) = \zeta_1(x), \quad q(x) = 0, \quad  r(x) = \zeta_2(x), \\
s(x) = \zeta_1(x)\zeta_2(x), \quad t(x) = \zeta_1^2(x), \quad u(x) = \zeta_2^2(x)/2,
\end{gather*}

\noindent where $\zeta_1(x)$ and $\zeta_2(x)$ are user-specified functions of $x$, subject to $\zeta_1(x_0) = 1$ and $\zeta_2(x_0) = 0$. This implies $\theta(x) = \mu(x) = \zeta_1(x)\mu_0 + \zeta_2(x)$.\\

As an example, choosing $\zeta_1(x) = 1$ and $\zeta_2(x) = x^T\beta$ gives rise to the random intercept model  $\mu(x) = \mu_{0} + x^T\beta$, where $x$ does not include the constant $1$ so that $\zeta_2(x_0) = 0$. Random slopes can be incorporated by writting $\mu(x)= z^T \mu_{0} + x^T \beta$, where $\mu_0$ is now a vector and $z$ is a known design matrix for the random effects (usually a subset of $x$).

\subsection{Poisson}
The Poisson density function can be written in the form
\begin{equation*}
\exp \left( y \log \mu_0 - \mu_0 - \log y!\right),
\end{equation*} 
where $\mu_0 > 0$ is the rate parameter. This can be written in the form of (\ref{eqn:expofam}) if we write $\theta_0 = \log \mu_0$, $b(\theta_0) = e^{\theta_0}$, $\phi = 1$ and $c(y,\phi) = - \log y!$. \\

To determine the conjugate distribution for $\theta_0$, we compute the normalization factor
\begin{equation*}
g(\chi,\nu) = \left\lbrace \int \exp\left(\chi\theta_0 - \nu \exp(\theta_0)\right) d\theta_0  \right\rbrace ^{-1} = \frac{\nu^\chi}{\Gamma(\chi)},
\end{equation*}
\noindent where the integrand is the kernel of a log-gamma density function with shape $A = \chi > 0$ and scale $B = \nu^{-1} > 0$, $\Gamma(\cdot)$ is the gamma function. This implies 
\begin{equation*}
\mu_0 = \exp(\theta_0) \sim \text{Gamma} \left( A, B  \right).
\end{equation*}\

\cite{Christiansen1997hierarchical} considered a similar model without covariates in the Bayesian setting.  Group level covariates can be incorporated via the mean of $\mu_0$, by letting E$(\mu_0) = AB \equiv \exp(x_i^T \beta)$ for example. As a result, we replace $B$ in the likelihood equation by $B_i = \exp(x_i^T \beta)/A$. To incorporate unit level covariates, (\ref{eqn:condition}) requires
\begin{equation*}
b(\theta(x)) = \exp(p(x)\theta_0 + q(x)\exp(\theta_0) + r(x)) \equiv s(x)\theta_0 + t(x)\exp(\theta_0) + u(x),
\end{equation*}
which gives the following solution set:
\begin{gather*}
p(x) = 1, \quad q(x) = 0, \quad  r(x) = \zeta(x), \\
s(x) = 0, \quad t(x) = e^{\zeta(x)}, \quad u(x) = 0,
\end{gather*}
where $\zeta(x)$ is a user-specified function of $x$, subject to $\zeta(x_0) = 0$. This implies $\theta(x) = \log(\mu(x)) = \theta_0 + \zeta(x)$, or equivalently, $\mu(x) = \mu_0 \exp(\zeta(x))$.\\

As an example, choosing $\zeta(x) = x^T\beta$ leads to $\mu(x) = \mu_0 \exp(x^T\beta)$, where $x$ does not include the constant $1$ so that $\zeta(x_0) = 0$. This is a sensible choice as $\mu(x)$ is guaranteed to be always positive. Similar multiplicative models with unit level covariates have been considered by  \cite{Lee2017sufficiency}, \cite{Lee2017poisson} and \cite{Lee1996} in various settings.

\subsection{Binomial (with known number of trials)}
The binomial density function (with fixed number of trials $n \in \mathbb{N}$) can be written in the form
\begin{equation*}
\exp \left\lbrace y \log\left(\frac{\mu_0}{1-\mu_0}\right) + n \log(1-\mu_0) + \log {n \choose y}  \right\rbrace,
\end{equation*} 
where $0 \leq \mu_0 \leq 1$ is the probability of success. This can be written in the form of (\ref{eqn:expofam}) if we write $\theta_0 = \log (\mu_0(1-\mu_0)^{-1})$, $b(\theta_0) = n \log\left( 1+\exp(\theta_0) \right)$, $\phi = 1$ and $c(y,\phi) = \log {n \choose y}$.\\

To determine the conjugate distribution for $\theta_0$, we compute the normalization factor
\begin{equation*}
g(\chi,\nu) = \left[ \int \exp\lbrace\chi\theta_0 - \nu \log\left(1+\exp(\theta_0)\right) \rbrace d\theta_0  \right] ^{-1} = \frac{1}{\text{B}(\chi,\nu-\chi)},
\end{equation*}
\noindent where the integrand is the kernel of the log of a beta prime density function with shape parameters $A = \chi > 0$ and scale $B = \nu-\chi > 0$, $\text{B}(\cdot)$ is the beta function. This implies 
\begin{equation*}
\mu_0 = \exp(\theta_0) / {(1+ \exp(\theta_0))} \sim \text{Beta} \left( A, B   \right).
\end{equation*}\

\cite{Kleinman1973proportions}, \cite{Crowder1978betabinomial} and \cite{He1998hierarchical} considered similar models without covariates in various settings. Group level covariates can be incorporated via the mean of $\mu_0$. Reparameterizing the beta density function by setting the mean $\lambda = A/(A+B)$ and precision $\phi = A+B$, we can allow $\lambda_i$ to be some function of $x_i^T \beta$, say, $\lambda_i = \lbrace 1+\exp(-x_i^T\beta) \rbrace^{-1}$ \cite[]{Ferrari2004}. As a result, we replace $A$ and $B$ in the likelihood equation by $\lambda_i\phi$ and $\phi-\lambda_i\phi$, respectively. To incorporate unit level covariates, (\ref{eqn:condition}) requires
\begin{gather*}
\hskip-40mm b(\theta(x)) = \log \left[ 1+ \exp \left\lbrace p(x)\theta_0 + q(x)\log(1+\exp(\theta_0)) + r(x) \right\rbrace \right] \equiv \\ \hskip80mm s(x)\theta_0 + t(x)\log\left(1+\exp(\theta_0)\right) + u(x),
\end{gather*}
which gives the following solution set:
\begin{gather*}
p(x) = 1, \quad q(x) = 0, \quad  r(x) = 0, \\
s(x) = 0, \quad t(x) = 1, \quad u(x) = 0.
\end{gather*}
Since neither $p$, $q$ nor $r$ depend on $x$, it is impossible to simultaneously incorporate unit level covariates while maintaining closed-form likelihood.

\subsection{Gamma (with known shape)}
For modelling purposes, it is convenient to reparameterize the gamma distribution with shape $A > 0$ and scale $B_0 > 0$ in terms of $A$ and mean $\mu_0 = AB_0 > 0$. The reparameterized gamma density function (with fixed shape $A$) can be written in the form
\begin{equation*}
\exp \left[  \frac{-y\mu_0^{-1} - \log \mu_0 }{A^{-1}} + A \log(Ay) - \log y - \log\Gamma (A) \right].
\end{equation*} 
This can be written in the form of (\ref{eqn:expofam}) if we write $\theta_0 = -\mu_0^{-1}$, $b(\theta_0) = -\log(-\theta_0)$, $\phi = A^{-1}$ and $c(y,\phi) =  A \log(Ay) - \log y - \log\Gamma (A)$.\\

To determine the conjugate distribution for $\eta_0$, we compute the normalization factor
\begin{equation*}
g(\chi,\nu) = \left[ \int \exp\left\lbrace \chi \theta_0- \nu (-\log(-\theta_0)) \right\rbrace d\theta_0  \right] ^{-1} = \frac{\chi^{\nu+1}}{\Gamma(\nu+1)},
\end{equation*}
\noindent where the integrand is the kernel of the negative of a gamma density function with shape $C = \nu+1$ and scale $D = \chi^{-1}$, and $\Gamma(\cdot)$ is the gamma function. This implies 
\begin{equation*}
\mu_0 = -\theta_0^{-1} \sim \text{Inverse-Gamma} \left( C , D  \right).
\end{equation*}\

Group level covariates can be incorporated via the mean of $\mu_0$, by letting E$(\mu_0) = D_i(C-1)^{-1} \equiv \exp(x_i^T \beta)$ for example, provided $C>1$. As a result, we replace $D$ in the likelihood equation by $D_i = (C-1)\exp(x_i^T \beta)$.  To incorporate unit level covariates, (\ref{eqn:condition}) requires
\begin{equation*}
b(\theta(x)) = -\log \lbrace p(x)\theta_0 + q(x) \log{(-\theta_0)} - r(x) \rbrace \equiv s(x)\theta_0 - t(x)\log{(-\theta_0)} + u(x),
\end{equation*}
which gives the following solution set:
\begin{gather*}
p(x) = \zeta(x), \quad q(x) = 0, \quad  r(x) = 0, \\
s(x) = 0, \quad t(x) = 1, \quad u(x) = -\log\zeta(x),
\end{gather*}
where $\zeta(x)$ is a user-specified function of $x$, subject to $\zeta(x_0) = 1$. This implies $\theta(x) = -\mu^{-1}(x) = \zeta(x)\theta_0$, or equivalently, $\mu(x) = \mu_0/\zeta(x)$.\\

As an example, choosing $\zeta(x) = \exp(x^T \beta)$ leads to $\mu(x) = \mu_0/\exp(x^T\beta)$, where $x$ does not include the constant 1 so that $\zeta(x_0) = 1$. This is a sensible choice as $\mu(x)$ is guaranteed to be always positive.

\subsection{Summary}
Table \ref{tab:group} and \ref{tab:unit} summarize the results discussed in this section, for group level and unit level models, respectively. We have covered the four distribution families that are most important in practice. Results for other distributions could be derived as needed.

\begin{table}
	\caption{Summary of group level models. Log-likelihood functions are contributed by a single observation. The $i$ and $j$ indexes are omitted for ease of notation.}
	\begin{tabular}{lc} 
		\multicolumn{2}{c}{\underline{Gaussian (with known variance $\sigma^2$)}}\\
		\rule{0pt}{3ex}    
		\hskip-1mm  Model & $y\mid\mu_{0} \sim \text{Gaussian}(\mu_{0},\sigma^2)$ \quad  $\mu_{0} \sim \text{Gaussian} (\lambda, \kappa^2)$ \\ 
		Log-likelihood & $-\frac{1}{2}  \left\lbrace \log(\sigma^2+\kappa^2) + \frac{y^2}{\sigma^2} + \frac{\lambda^2}{\kappa^2} - \frac{\lambda^2\sigma^4 + 2 \lambda\kappa^2\sigma^2 y + \kappa^4  y^2}{\kappa^2\sigma^2(\sigma^2+\kappa^2)}\right\rbrace$ \\ 
		Covariates & Replace $\lambda$ by $\lambda_i = x_i^T \beta$ \\ 
		\\ 
		\multicolumn{2}{c}{\underline{Poisson}} \\
		\rule{0pt}{3ex}   
		\hskip-1mm  Model & $y\mid\mu_{0} \sim \text{Poisson}(\mu_{0})$ \quad  $\mu_{0} \sim \text{Gamma} (A, B)$ \\ 
		Log-likelihood & $  \log \Gamma\left(A+ y\right) - \left(A+y\right)\log(B^{-1}+1) - \log\Gamma(A) - A \log B  $ \\ 
		Covariates & Replace $B$ by $B_i = e^{x_i^T \beta} / A$ \\ 
		\\
		\multicolumn{2}{c}{\underline{Binomial (with known number of trials $n$)}} \\
		\rule{0pt}{3ex}   
		\hskip-1mm  Model & $y\mid \mu_{0} \sim \text{Bernoulli}(\mu_{0})$ \quad  $\mu_{0} \sim \text{Beta} (A,B)$ \\ 
		Log-likelihood &  $ \log B\left(A+ y, B+1-y\right) - \log B(A,B) $ \\ 
		Covariates & Replace $A$ and $B$ by $\lambda_i \phi$ and $\phi-\lambda_i\phi$ respectively, where $\lambda_i = \left\lbrace 1+ e^{-x_i^T \beta} \right\rbrace ^{-1}$ \\ 
		\\
		\multicolumn{2}{c}{\underline{Gamma (with known shape $A$)}} \\
		\rule{0pt}{3ex}   			
		\hskip-1mm  Model & $y\mid\mu_{0} \sim \text{Gamma}(A, \mu_{0}/A)$ \quad  $\mu_{0} \sim \text{Inverse-Gamma} (C, D)$ \\ 
		Log-likelihood & $ - \log B(A,C) + A \log (ADy) -\log y + (A+C) \log(1+ADy)  $ \\
		Covariates & Replace $D$ by $D_i = (C-1)e^{x_i^T \beta}$, provided $C>1$ 
	\end{tabular} \label{tab:group}
\end{table}

\begin{table}
	\caption{Summary of unit level models. Log-likelihood functions are contributed by a single observation. The $i$ and $j$ indexes are omitted for ease of notation.}
		\begin{tabular}{lc}
			\\
			\multicolumn{2}{c}{\underline{Gaussian (with known variance $\sigma^2$)}}\\
			\rule{0pt}{3ex} 
			\hskip-1mm Model & $y\mid\mu_{0} \sim \text{Gaussian}\left(\zeta_1(x)\mu_{0} +  \zeta_2(x),\sigma^2\right)$ \quad  $\mu_{0} \sim \text{Gaussian} (\lambda, \kappa^2)$ \\ 
			Constraint & $\zeta_1(x_0) = 1$ \quad $\zeta_2(x_0)=0$ \\
			Log-likelihood & $-\frac{1}{2} \left\lbrace \log\left(\sigma^2+\kappa^2\sum_j\zeta_1^2(x)\right) + \frac{\sum_j y^2}{\sigma^2} + \frac{\lambda^2}{\kappa^2}  - \frac{2\sum_j\zeta_2(x)y}{\sigma^2} + \frac{\sum_j\zeta_2^2(x)}{\sigma^2} + \frac{P}{Q} \right\rbrace$ \\ 
			& where $P = -\lambda^2\sigma^4 + 2\lambda\kappa^2\sigma^2\left(\sum_j\zeta_1(x)y\right) - \kappa^4\left(\sum_j\zeta_1(x)y\right)^2 +  $ \\
			&  $ 2\kappa^4\left(\sum_j\zeta_1(x)y\right)\left(\sum_j\zeta_1(x)\zeta_2(x)\right) - \kappa^4\left(\sum_j\zeta_1(x)\zeta_2(x)\right)^2 - $ \\
			& $ 2\lambda\kappa^2\sigma^2\left(\sum_j\zeta_1(x)\zeta_2(x)\right)$ \\
			&  $Q = \kappa^2\sigma^2\left(\sigma^2+\kappa^2\sum_j\zeta_1^2(x)\right)$\\
			Remark & Can incorporate random slopes if $\mu(x)$ is linear in terms of $\mu_0$\\
			\\
			\multicolumn{2}{c}{\underline{Poisson}}\\
			\rule{0pt}{3ex} 
			\hskip-1mm  Model & $y\mid\mu_{0} \sim \text{Poisson}\left(\mu_{0}e^{\zeta(x)}\right)$ \quad  $\mu_{0} \sim \text{Gamma} (A, B)$ \\ 
			Constraint & $\zeta(x_0) = 0$ \\
			Log-likelihood & $ \log \Gamma\left(A+\sum_j y\right) - \left(A+\sum_j y\right)\log\left(B^{-1}+\sum_j e^{\zeta(x)}\right) - $\\
			& $\log\Gamma(A) -A \log B  + \sum_j\zeta(x)y$ \\ 
			\\
			\multicolumn{2}{c}{\underline{Gamma (with known shape $A$)}}\\
			\rule{0pt}{3ex} 
			\hskip-1mm  Model & $y\mid\mu_{0} \sim \text{Gamma}(A, \mu(x)/A)$ \quad  $\mu_{0} \sim \text{Inverse-Gamma} (C, D)$ \\ 
			& where $\mu(x) = \mu_{0}/\zeta(x)$ \\
			Constraint & $\zeta(x_0) = 1$ \\
			Log-likelihood & $\log \Gamma (An_i+C) - n_i \log\Gamma(A) - \log\Gamma(C) + An_i \log A + (A-1) \sum_j\log y  -$ \\ 
			& $  (An_i+C)\log\left\lbrace1+ AD\left(\sum_j \zeta(x)y\right) \right\rbrace + An_i \log D + A \sum_j \log \zeta(x)$ \\
			&  where $n_i$ is the number of units within group $i$
		\end{tabular}	\label{tab:unit}
\end{table}

\section{An Illustrative Example: Poisson responses}
Consider the well-known epileptic seizure count data previously analyzed by \cite{Thall1990}, \cite{Breslow1993}, \cite{Lee1996} and \cite{Ma2007}, where $59$ epileptics were randomized to a new drug (Trt = progabide) or a placebo (Trt = placebo) at a clinical trial. Baseline data included the log seizure counts during the 8-week period before the trial (lbase) and the log age in years (lage), both centered to have zero mean. A multivariate response variable consisted of the counts seizures during the 2-weeks before each of four clinic visits. An indicator variable for the fourth visit (V4) was constructed to reflect the fact that counts are substantially lower during the fourth visit. The dataset are stored in the epil object within the MASS package in \textsf{R} \cite[]{R2017}. \\

Our reanalysis is primarily oriented toward comparing two different methods of incorporating random effects, namely, generalized linear mixed models (GLMM) using additive Gaussian random effects: $y_{ij}|u_i \sim \text{Poisson}(\exp(x_{ij}^T\beta+u_i))$,   $\mu_i \sim \text{Gaussian} (0,\sigma^2)$; and conjugate generalized linear mixed models (CGLMM) using multiplicative Gamma random effects: $y_{ij}|u_i \sim \text{Poisson}(u_i \exp(x_{ij}^T\beta))$, $\mu_i \sim \text{Gamma} (A,1/A)$. To allow for a direct comparison between the models, we included an intercept in the Poisson conjugate generalized linear mixed model, but fixed the  mean of $u_i$ to be one to ensure identifiability. Due to the intractable nature of the  marginal likelihood of Poisson generalized linear mixed models, various approximation methods have been employed to estimate the marginal likelihood. The results are presented in Table \ref{tab:results}.\\

\begin{table}[h]
	\centering
	\caption{Regression estimates for the epileptics data, and the associated standard errors.}
	\label{tab:results}
	\begin{threeparttable}
		\resizebox{\textwidth}{!}{%
			\begin{tabular}{lccccc}          
				\toprule\midrule
				&   &  \multicolumn{3}{c}{\textbf{GLMM}} \\
				\cline{3-5}
				\textbf{Variables}  & \textbf{GLM} & \textbf{Laplace$^1$} & \textbf{AGQ$^2$}& \textbf{PQL$^3$} & \textbf{CGLMM} \\
				\midrule
				Intercept     & 1.898 (0.043)   & 1.833 (0.105)    & 1.833 (0.106)    & 1.870 (0.106)  & 1.932 (0.105) \\
				lbase     & 0.949 (0.044)   & 0.883 (0.131)    & 0.883 (0.131)    & 0.882 (0.129)  & 0.880 (0.126) \\
				trtprogabide    & -0.346  (0.061)  & -0.334 (0.147)    & -0.334 (0.148)    & -0.310 (0.149)  & -0.282 (0.146) \\
				lage    & 0.888 (0.116)   & 0.481 (0.346)    & 0.481 (0.347)    & 0.534 (0.346)  & 0.505 (0.357) \\
				V4      & -0.160 (0.055) & -0.160 (0.054)  & -0.160 (0.055)  & -0.160 (0.077) & -0.160 (0.055) \\
				lbase:trtprogabide  & 0.562 (0.064)             & 0.339 (0.202) & 0.339   (0.203)           & 0.342  (0.203)            & 0.344 (0.193)  \\
				$\sigma$  & N/A             & 0.501 (N/A)             & 0.502   (N/A)           & 0.444 (N/A)  &  N/A\\
				$A$  & N/A             & N/A              & N/A              & N/A &  3.935 (0.863) \\
				\bottomrule\addlinespace[1ex]
			\end{tabular}}
			\begin{tablenotes}\footnotesize
				\item[$^1$] Laplace approximation: fitted using the glmer() function within the lme4 package in \textsf{R}.
				\item[$^2$] Adaptive Gauss-Hermite quadrature: fitted using the glmer() function within the lme4 package in \textsf{R}, \\ using nAGQ=100.
				\item[$^3$] Penalized Quasi-Likelihood: fitted using the glmmPQL() function within the MASS package in \textsf{R}.
			\end{tablenotes}
		\end{threeparttable}
	\end{table} 
	
	In comparing the estimates and standard errors between the models, we note that the fixed effects model is likely to produce biased estimates as it did not take into account for the correlation induced by multiple measurements from the same individual. The parameter estimates and the standard errors of the random effect models are quite similar, regardless of the distribution of the random effects. This is probably due to the fact that the variance of the random effects not being too large, implying moderate subject-to-subject variability in seizure counts after taking into account of the covariate effects.


\section{Remarks}

Group level conjugate models have long been used in the context of Bayesian small area estimation and disease mapping, the most common ones being the gamma-Poisson \cite[p. 383]{Rao2015} and the beta-binomial models \cite[p. 389]{Rao2015}. This article considers the frequentist setting where the most general conditions that allow for explicit marginal likelihood in unit level generalized linear mixed models are derived. The primary advantage of the proposed modelling framework is mathematical convenience, but the conjugate random effect distribution this assumes may not accurately reflect the real variation between groups. Mathematical convenience should not deter the exploration of alternative formulations for the distribution of random effects in this situation. Other applications of the proposed modelling framework include privacy preservation in large-scale administrative databases \cite[]{Lee2017sufficiency} and the fitting of discrete choice models \cite[]{Lee2017poisson}. \\

Some of the models derived from our conjugate generalized linear mixed models framework are similar to those of the \textit{conjugate hierarchical generalized linear models} framework proposed by \cite{Lee1996}. While the word ``conjugate'' in our framework reflects the fact that the marginal likelihood can be made explicit, it has quite a different meaning in the hierarchical likelihood framework \cite[p. 621]{Lee1996}, where it refers to the fact that a Bayesian conjugate prior is imposed on the random effects distribution but does not necessarily result in a closed-form likelihood. \\

\cite{Molenberghs2010} considered models that can simultaneously accommodate both overdispersion and correlation induced by grouping structures via two separate sets of random effects. They consider a combined model where the conjugate and Gaussian random effects induce overdispersion and association, respectively. Although they use the conjugate distribution for a set of random effects, the resulting marginal likelihood is generally not explicit.

\section*{Acknowledgements}
Lee's research is partially supported by the Australian Bureau of Statistics.


\newpage

\bibliographystyle{agsm} 

\bibliography{paper-ref} 

\end{document}